\documentclass[aps,prl,reprint,superscriptaddress,amsmath,amssymb,showpacs,floatfix]{revtex4-1}
\usepackage{graphicx}
\usepackage{bm}
\usepackage{color}

\newcommand{\ndcoeff}{\varepsilon} 

\begin{document}

\title{Multiple dynamic transitions in nonequilibrium work fluctuations}
\author{Jae Dong Noh}
\affiliation{Department of Physics, University of Seoul, Seoul 130-743,
 Korea}
\affiliation{School of Physics, Korea Institute for Advanced Study,
Seoul 130-722,  Korea}

\author{Chulan Kwon}
\affiliation{Department of Physics, Myongji University, Yongin, Gyeonggi-Do,
449-728,  Korea}

\author{Hyunggyu Park}
\affiliation{School of Physics, Korea Institute for Advanced Study,
Seoul 130-722, Korea}

\date{Februrary 23, 2012}

\begin{abstract}
The time-dependent work probability distribution function $P(W)$ is
investigated analytically for a diffusing particle trapped by an anisotropic
harmonic potential and driven by a nonconservative drift force in two
dimensions.
We find that the exponential tail shape of $P(W)$ characterizing rare-event
probabilities undergoes a sequence of dynamic transitions in time. These remarkable {\em locking-unlocking} type
transitions result
from an intricate interplay between a rotational mode induced by the nonconservative
force and an anisotropic decaying mode due to the conservative attractive force. We expect
that most of high-dimensional dynamical systems should exhibit similar multiple dynamic transitions.

\end{abstract}
\maketitle

Systems in thermal equilibrium are governed by the principle of statistical
mechanics. It serves as the unified framework for the study of thermodynamic
properties of equilibrium systems, and has been
successful since it was established centuries ago.
On the contrary, such a principle, except for the second law of thermodynamics,
was absent for nonequilibrium systems, which made it difficult to understand
nonequilibrium phenomena. Recently, discovery of the fluctuation theorem
opened a new perspective on nonequilibrium processes and has attracted a lot of
interests~\cite{Seifert12}.
The fluctuation theorem refers to identity relations for a thermodynamic quantity, such
as work, heat, or entropy production, that are derived theoretically for a wide class
of nonequilibrium process~\cite{Evans93,Gallavotti,Jarzynski97,
Crooks99,Lebowitz99,Hatano01,Seifert05,Kurchan07,Sevick08,Noh12}, and are also confirmed
experimentally~\cite{Carberry04,Reid04,Collin05,Garnier05,Ciliberto10,Hayashi10}.
It not only serves as a criterion diagnostic to nonequilibrium, but also
sheds light on  quantitative understanding of nonequilibrium fluctuations~\cite{Harada05,Seifert06,Blickle07,Prost09,Mallick11}.


The  fluctuation theorem evokes the importance of studying nonequilibrium fluctuations,
especially in the rare-event region.
Many studies have been done for the probability distribution
functions~(PDF's) of the work and heat
associated with nonequilibrium processes, theoretically and experimentally~\cite{Evans93,Gallavotti,Jarzynski97,
Crooks99,Lebowitz99,Hatano01,Seifert05,Kurchan07,Sevick08,Noh12,Carberry04,Reid04,Collin05,
Garnier05,Ciliberto10,Kwon11,Filliger07,JS02,Evans02,Zon03,Ciliberto,Ciliberto13}.
These are to confirm the fluctuation theorem in the
first place, and then to investigate the nature of nonequilibrium
fluctuations. Normally it is a formidable task to find the PDF analytically
for a specific nonequilibrium process, so most studies
are limited to special cases such as the large deviation study
in the infinite-time limit~\cite{Farago02,Fogedby11,Visco06,Puglisi06}.

In this Letter, we investigate the PDF of a nonequilibrium work $W$
over a {\em finite} time interval $t$ in a two-dimensional linear diffusion
system (LDS), driven by a nonconservative linear drift  force.
The LDS is often referred to as a multivariate Ornstein-Uhlenbeck process
describing the motion of a Brownian particle trapped
by a linear force in the overdamped limit~\cite{Gardiner,Risken}.
It is also well known that the LDS serves as a generic model to describe
the fluctuation effects on deterministic dynamic solutions of general nonlinear
systems in the context of the van Kampen's system size expansion~\cite{Gardiner}.
Experimental systems described by the LDS are so diverse, including
a nano heat engine in contact
with multiple heat reservoirs~\cite{Filliger07},
a colloidal particle driven along periodic potential
imposed by laser traps~\cite{Blickle07}, biological molecular motor
systems~\cite{Hayashi10}, electric circuits~\cite{Ciliberto13},
global climate systems~\cite{Weiss}, and musical instruments~\cite{Gannot}.

The LDS is simple enough that the PDF $P(W)$ is analytically tractable
for finite time interval.
Our study reveals that
even such a simple system displays a surprisingly rich dynamic behavior
with a sequence of dynamic transitions in time.  We briefly
summarize the results with dimensionless $W$:
(i) The PDF has exponential tails with power-law
prefactors as $P(W) \sim |W|^{-\alpha_\pm }\ e^{-W/W_{\pm}}$ for large  $|W|$.
The power-law exponents are the same in both
sides~($\alpha_+=\alpha_-=\alpha$),
and the characteristic works $W_+>0$ and $W_-<0$ satisfy
$1/W_+ + 1/W_- = -1$, which are required by the detailed fluctuation theorem~\cite{Crooks99,Reid04,Kwon11}.
(ii) The power-law exponent $\alpha$ can take  three different values of $0$,
$1/2$, and $2$. Accordingly, the PDF is categorized into type 0 with
$\alpha = 0$, type I with $\alpha=1/2$, and type II with $\alpha=2$.
Interestingly, $W_{\pm}$ continuously varies with time $t$ for type 0 and type I,
while they are constants of time for type II.
(iii) Typically, the system undergoes a dynamic transition
from type I to type II as $t$ increases. The characteristic work $W_+$
increases smoothly with $t$ and suddenly becomes frozen
at the transition time $t_c$ and afterwards. This is a kind of a
{\em locking} transition.
More remarkably, in some parameter space, the PDF alternates between type I and type II
indefinitely, i.e.~infinite number of locking-unlocking type transitions.
In general, a finite sequence of dynamic
transitions is also possible as well as no transition with type I in all time.
Type 0 is found without any dynamic transition, only in a special case.

We consider a LDS with the equations of motion
\begin{equation}\label{eq_motion_tr}
\frac{d{\bm q}}{dt} = -{\mathsf F} \cdot {\bm q} + {\bm \xi}
\end{equation}
where ${\bm q} = ({q}_1,\cdots,{q}_d)^T$ is a
$d$-dimensional vector, ${\mathsf F} = ({F}_{ij})$
is a constant positive-definite $(d\times d)$ force matrix,
and ${\bm \xi}(t)=({\xi}_1(t),\cdots,{\xi}_d(t))^T$
is the Langevin noise satisfying
\begin{equation}
\langle {\xi}_i(t) \rangle = 0 \ , \
\langle {\xi}_i(t) {\xi}_j (t)\rangle =
2 {D}_{ij} \delta(t-t')
\end{equation}
with a noise correlation matrix $\mathsf{D}= ({D}_{ij})$
which is symmetric and positive-definite.
After a similarity and a scale transformation, one can take
the noise matrix as the identity matrix $\mathsf{I}$ ($\mathsf{D} = \mathsf{I}$)
without loss of generality.


The force matrix can be decomposed into the conservative and nonconservative parts as $\mathsf{F}
= {\mathsf F}_c + {\mathsf F}_{nc}$ with symmetric ${\mathsf F}_c=\mathsf{F}_c^T$
and nonsymmetric ${\mathsf F}_{nc}$.
When ${\mathsf F}_{nc}=0$,
the total force ${\bm f} = -\mathsf{F}\cdot {\bm q}=  - {\mathsf \nabla} V({\bm q})$ is conservative
with a potential function
$V({\bm q}) = \frac{1}{2}{\bm q}^T \cdot \mathsf{F} \cdot {\bm q}$.
The steady-state PDF is given by the equilibrium
Boltzmann distribution $P_{eq}({\bm q}) \propto e^{-V({\bm q})}$.
A nonsymmetric force
matrix~(${\mathsf{F}}_{nc} \neq \mathsf{0}$)
indicates the presence of a nonconservative force
$\bm{f}_{nc} = -{\mathsf{F}}_{nc} \cdot \bm{q}$ which cannot be
written as a gradient function. It drives the system out of equilibrium.
Here, we only consider the antisymmetric ${\mathsf{F}}_{nc}$ ($=-{\mathsf{F}}_{nc}^T$)
for simplicity~\cite{exp}.

Suppose that the system is prepared in the thermal equilibrium with
$P_{eq}({\bm q}) \propto e^{-({1/2}){\bf q}^T \cdot {\mathsf F}_c \cdot {\bm
q}}$  with the conservative force $\bm{f}_{c} = -{\mathsf F}_c \cdot \bm{q}$
only,  then turn on the nonconservative force
$\bm{f}_{nc}$ at $t=0$. The nonequilibrium
work (done by $\bm{f}_{nc}$) on the particle following a path
${\bm q}(\tau)$ for $0\leq \tau\leq t$ is given by
\begin{equation}\label{W_def}
\mathcal{W}[{\bm q}(\tau)] =
-\int_0^t d\tau \ \frac{d{\bm q}(\tau)^T}{d\tau} \cdot {\mathsf F}_{nc}
\cdot \bm{q}(\tau) \ .
\end{equation}
We are interested in the PDF $P(W) = \langle
\delta(W-\mathcal{W}[{\bm q(\tau})] )\rangle$ and
its characteristic function
$\mathcal{G}(t;\lambda) = \langle e^{-\lambda W} \rangle =\int dW e^{-\lambda W} P(W)$,
which should satisfy the fluctuation theorem symmetry as $P(W)/P(-W)=e^W$ and consequently
${\mathcal G}(\lambda)={\mathcal G}(1-\lambda)$~\cite{Kwon11}.

The characteristic function can be written as a path integral
${\mathcal G}(\lambda) \propto \int D[{\bm q}] e^{-L[{\bm
q};\lambda]}$ with an action $L$. In our previous work~\cite{Kwon11}, we developed a
formalism evaluating the path integral for the LDS where the
action $L$ is quadratic in ${\bm q}$.
The key idea is described as below.
The Gaussian integration can be performed
successively from ${\bm q}(0)$ to ${\bm q}(t)$ at discretized times.
Integration up to ${\bm q}(\tau)$ yields a
modified kernel for ${\bm q}(\tau+\Delta t)$, denoted by a symmetric
$d\times d$ matrix $\widetilde{\mathsf A}(\tau+\Delta \tau;\lambda)$.
Comparing the kernels at $\tau$ and $\tau+\Delta \tau$ and taking the limit
$\Delta \tau\to0$, one can derive the differential equation for the kernel $\widetilde{\mathsf A}(\tau)$ as
\begin{equation}\label{dAdt}
\frac{d}{d\tau}\tilde{\mathsf A}(\tau;\lambda) =
- 2 \tilde{\mathsf A}^2 + \tilde{\mathsf A}
\tilde{\mathsf F} + \tilde{\mathsf F}^T \tilde{\mathsf{A}} + \bm{\Lambda}
\end{equation}
with the initial condition $\tilde{\mathsf A}(0) = {\mathsf F}_c$ and
the auxiliary matrices
$\tilde{\mathsf{F}}(\lambda) = \mathsf{F} - 2\lambda{\mathsf{F}}_{nc}$ and
$\bm{\Lambda}(\lambda) = (\mathsf{F}^T \mathsf{F} - \tilde{\mathsf{F}}^T
\tilde{\mathsf{F}})/2$. The characteristic function is then given by the product of
Gaussian integrals with the kernel $\widetilde{\mathsf A}(\tau)$ along the
path ${\bm q}(\tau)$, which yields
\begin{equation}\label{lnG}
\ln \mathcal{G}(t;\lambda) = - \int_0^t d\tau \ {\rm Tr} (
\tilde{\mathsf{A}}(\tau;\lambda) - \tilde{\mathsf{F}})
-\frac{1}{2} \ln \frac{ \det \tilde{\mathsf{A}}(t;\lambda)}{
\det {\mathsf{F}}_c} .
\end{equation}


We apply the formalism to a two-dimensional system with
the force matrices, parameterized as~\cite{comment}
\begin{equation}
{\mathsf F}_c = \left( \begin{array}{cc}
1+u & 0 \\ 0 & 1-u \end{array} \right) , \
\mathsf{F}_{nc} =
\left( \begin{array}{cc} o &\  \ndcoeff \\
-\ndcoeff &\  0 \end{array} \right) \ .
\end{equation}
One can set the trace of $\textsf{F}$ to any positive number
by the global rescaling of ${\bm q}$ and $t$.
Here, it is set to be $2$.
Positive-definiteness of $\mathsf{F}$ and ${\mathsf F}_c$ requires that
$u^2<1$.
The system describes a Brownian particle trapped by
an anisotropic harmonic potential and driven by a rotational torque.
The parameter $\ndcoeff$ corresponds to the strength of
nonequilibrium driving (torque), while the parameter $u$ represents the
anisotropy of the harmonic potential.

Before considering the general case, we present the
solution in the special isotropic case with $u=0$. In this case, the matrix
$\tilde{\mathsf A}$ is proportional to the identity matrix
as $\tilde{\mathsf A}=z(\tau;\lambda)\ {\mathsf I}$. Then, Eq.~(\ref{dAdt})
becomes ${dz}/{d\tau} = -2z^2 + 2z + 2\ndcoeff^2 \lambda(1-\lambda)$ with
$z(0)=1$ and Eq.~(\ref{lnG}) is written as
\begin{equation}\label{lnG0}
\ln \mathcal{G}(t;\lambda) =  -2\int_0^{t} (z(\tau;\lambda)-1)d\tau -\ln
z(t;\lambda)\ .
\end{equation}
The solution of the differential equation is given by
\begin{equation}\label{z_s}
z(\tau;\lambda) = \frac{1}{2}\left( 1 + \sqrt{\Delta} \frac{ 1 + \sqrt{\Delta}
\tanh(\sqrt{\Delta}\tau)}{ \sqrt{\Delta} + \tanh(\sqrt{\Delta}\tau)} \right)
\end{equation}
with $\Delta(\lambda) \equiv 1 - 4 \ndcoeff^2 \lambda(\lambda-1)$.
Inserting Eq.~(\ref{z_s}) into Eq.~(\ref{lnG0}), we obtain ${\mathcal
G}(t;\lambda)$ and draw the following conclusions: (i)~Since
$\Delta(\lambda)=\Delta(1-\lambda)$, one finds $\mathcal{G}(t;\lambda) =
\mathcal{G}(t;1-\lambda)$ at all $t$. This verifies the fluctuation theorem.
(ii)~Given $t$, there exists $\lambda_0(t) > 1$ such that
$z(t;\lambda=\lambda_0) = z(t;\lambda=1-\lambda_0) = 0$.
The logarithmic term in Eq.~(\ref{lnG0}) indicates simple poles
of $\mathcal{G}(t;\lambda)$ at $\lambda=\lambda_0$ and $1-\lambda_0$.
(iii)~The simple poles manifest the exponential tails
of $P(W)$:
\begin{equation}
P(W) \sim \left\{ \begin{array}{ll}
e^{-(\lambda_0(t)-1)W} &,\  W \to \infty \\ [2mm]
e^{ -\lambda_0(t) |W|} &,\  W \to -\infty
\end{array} \right.\label{NP}
\end{equation}
with $\lambda_0(t)$ monotonically decreasing in $t$
and asymptotically approaching
$\lambda_c= \frac{1}{2}+{\sqrt{1+\ndcoeff^2}}/{(2\ndcoeff)}\ge 1$.
This PDF belongs to type 0.

The analysis above gives us a lesson that the singularity of $\mathcal{G}$,
hence the tail behavior of $P(W)$, is determined from the root of $\det
\tilde{\mathsf{A}}(t;\lambda)=0$. In the  isotropic potential case ($u=0$), two
eigenvalues of $\tilde{\mathsf A}$ are degenerate. So, the root contributes
to a simple pole of $\mathcal{G}$ and the pure exponential tail of $P(W)$.
When the degeneracy is broken, which is the case for an anisotropic
potential with $u\neq 0$, $\mathcal{G}$ has a square-root
singularity and one may simply expect the PDF of type I.
However, the actual behavior turns out to be much richer.
The particle driven by the torque undergoes an energy barrier periodic in the azimuthal direction,
due to the anisotropy  in the potential.
Therefore there emerges an extra time scale to overcome the barrier
in addition to the overall relaxation time.
Their interplay can be understood from the full solution of Eq.~(\ref{dAdt}).

Here, we sketch briefly the way to find the exact solution of
Eq.~(\ref{dAdt}). First, note that the inhomogeneous quadratic differential
equation, Eq.~(\ref{dAdt}), can be transformed into a solvable homogeneous linear
differential equation by shifting and inverting $\mathsf{\tilde A}(\tau;\lambda)$
such as
\begin{equation}\label{A_solution}
\mathsf{\tilde A}(\tau;\lambda) \equiv \mathsf{\tilde A}_s(\lambda) +
\mathsf{R}(\tau;\lambda)^{-1} \ ,
\end{equation}
where $\tilde{\mathsf A}_s(\lambda)$ is the fixed-point solution
satisfying $d\tilde{\mathsf A}/{d\tau}|_{\tilde{\mathsf A}=\tilde{\mathsf A}_s}=\mathsf{0}$.
Then, it is straightforward to derive
\begin{equation}\label{dRdt}
\frac{d\mathsf{R}}{d\tau} =
2 \mathsf{I} - \mathsf{R \hat{F}^T - \hat{F} R}  \ ,
\end{equation}
where $\hat{\mathsf F} (\lambda) \equiv \tilde{\mathsf F} (\lambda)- 2 \tilde{\mathsf A}_s(\lambda)$.
Its solution is given as
\begin{equation}
\mathsf{R}(\tau;\lambda) = e^{-\tau \hat{\mathsf F}}\ \mathsf{R}(0;\lambda)\
e^{-\tau\hat{\mathsf F}^T} + 2 \int_0^\tau d\tau' e^{-\tau' \hat{\mathsf F}}
e^{-\tau' \hat{\mathsf F}^T} .
\end{equation}
In two dimensions, the explicit expression for
Eq.~(\ref{A_solution}) is available. It is rather complex, and will be
presented elsewhere~\cite{Noh11}.

\begin{figure}
\centering
\includegraphics*[width=\columnwidth]{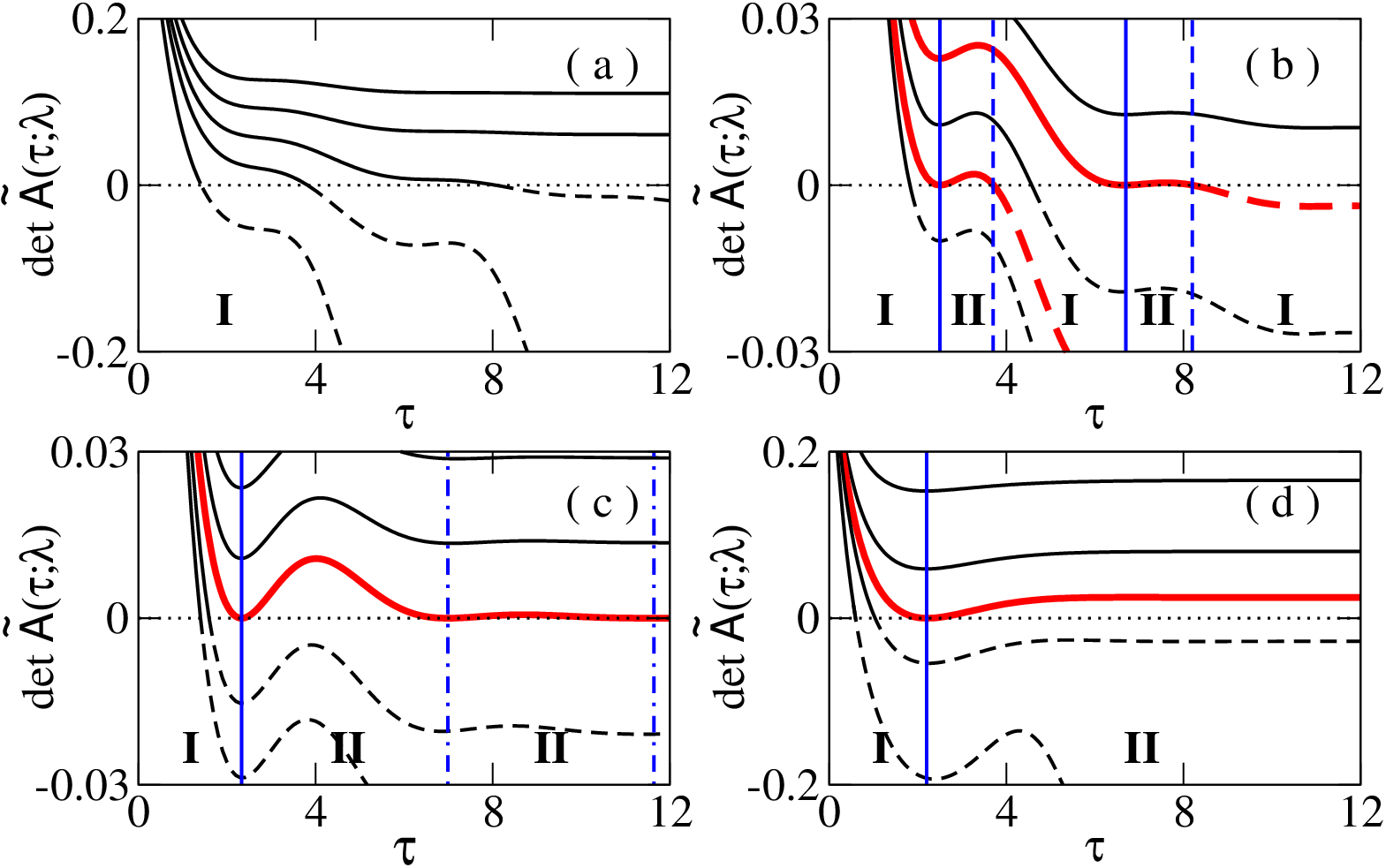}
\caption{(Color online)
Time evolution of $\det\tilde{\mathsf A}(\tau;\lambda)$ when
$\ndcoeff=3/10$ and (a) $u=0.65$, (b) $0.7$, (c) $u^* = \sqrt{109/200}$,
and (d) $0.8$ with several values of $\lambda>1$~(lower curves represent larger
values of $\lambda$). The curves tangential to the
$\tau$ axis~(dotted horizontal line) are drawn with thick
lines~(red). The characteristic function $\mathcal{G}$ is well defined only
when $\det\tilde{\mathsf A}>0$. Curves in the unphysical
regions with $\det\tilde{\mathsf A}<0$ are plotted with dashed lines.}
\label{fig1}
\end{figure}

The solution is the starting point for further analysis of the PDF.
We found that $\det{\tilde{\mathsf A}(t;\lambda)}$ exhibits
a complex behavior depending on values of $\ndcoeff$ and $u$.
There are four distinct cases, which are shown in Fig.~\ref{fig1}.
The plots are obtained for a few values of $u$ to a given $\ndcoeff=3/10$.

(i) When $u$ (anisotropy) is small enough~(see Fig.~\ref{fig1}(a)),
the curve is not tangential to the $\tau$ axis at any value of $\lambda$.
For any given $\tau=t$, one can find $\lambda_0$ such that
$\det\tilde{\mathsf A}(\tau,\lambda_0)=0$ (non-degenerate root).
Then, $\det\tilde{\mathsf A}(t,\lambda) \sim (\lambda_0-\lambda)$
near $\lambda\simeq \lambda_0$ and, from Eq.~(\ref{lnG}),
$\mathcal{G}(t,\lambda) \sim (\lambda_0-\lambda)^{-1/2}$.
Hence, the PDF has a tail $P(W)\sim |W|^{-1/2}
e^{\lambda_0 W}$ in the $W\to -\infty$ limit~(type I), with
$\lambda_0=\lambda_0(t)$ monotonically decreasing with $t$ to
an asymptotic value $\lambda_c(u,\ndcoeff)>1$.

(ii) In the intermediate values of $u$~(see Fig.~\ref{fig1}(b)), the curve
is tangential to the $\tau$ axis at multiple values of $\lambda$.
We will denote the time for the $n$-th tangential point as
$t_n^+$~(marked with vertical solid lines)
and the corresponding $\lambda$ value as $\lambda_n$. The curve,
$\det\tilde{\mathsf A}(\tau,\lambda_n)$, that
is tangential at $\tau=t_n^+$ may cross the $\tau$ axis
at a later time denoted as $\tau=t_n^-$~(marked with vertical dashed
lines). This crossing is linear (non-degenerate root) and never happens
again later. Within the time interval
$t_n^+ < t < t_n^-$, the characteristic function $\mathcal{G}(t;\lambda)$ is
finite for $\lambda < \lambda_n$, and then it diverges discontinuously at
$\lambda=\lambda_n$. Hence the PDF has a tail $P(W) \sim |W|^{-2}
e^{\lambda_n W}$ in the $W \to -\infty$ limit~(type II)~\cite{exp2}.
Note that $\lambda_n$ is a constant of time within the finite time interval.
Outside the interval, the PDF belongs to type I. Hence, the PDF
alternates between type I and type II many times as $t$ increases.

(iii) At the special value of $u=u^* \equiv \sqrt{(1+\ndcoeff^2)/2}$,
the curve, $\det\tilde{\mathsf A}(\tau,\lambda^*)$, with
$\lambda^* = (1+\ndcoeff)/(2\ndcoeff)$
is tangential to the $\tau$ axis infinitely many times as
\begin{equation}
\det\tilde{\mathsf A}(\tau;\lambda^*) =
\frac{ \ndcoeff(1-\ndcoeff)(1+\ndcoeff) \left( 1 + \cos \omega \tau \right)}{
4 \left( e^{2\ndcoeff\tau} - \frac{1-\ndcoeff}{1+\ndcoeff} -
\ndcoeff\frac{1-\ndcoeff}{1+\ndcoeff} \cos\omega \tau \right)} \ ,
\end{equation}
with $\omega = \sqrt{2(1-\ndcoeff^2)}$ and the tangential points at
$t_n^+ = (2n-1)\pi/\omega$ $(n=1,2,\ldots)$. Note that $t_n^-
= t_{n+1}^+$. Hence, the PDF belongs to type I when $t<t_1^+$~(marked
with the vertical solid line) and changes to type II afterwards
except periodic instantaneous moments at $t= t_n^+$ ($n\ge2 $)~(marked with
vertical dot-dashed lines).

(iv) When $u>u^*$, the curve is tangent to the $\tau$ axis at a
single value of $\lambda$ at $\tau = t_1^+$~(marked with the vertical solid
line) without crossing the $\tau$ axis later.
Hence, the PDF belongs to type I for
$\tau<t_1^+$ and type II afterwards forever.

Our analysis reveals that the system undergoes a
dynamic transition in the tail shape of the PDF between type I
characterized by $P(W) \sim |W|^{-1/2} e^{\lambda_0 W}$
and the type II characterized by
$P(W) \sim |W|^{-2} e^{\lambda_0W}$ for large negative $W$. The same applies
for large positive $W$ due to the fluctuation theorem symmetry.
The parameter $\lambda_0$, which corresponds to the inverse of the characteristic
fluctuation size $|W_-|$ for the negative PDF tail,
decreases smoothly in time for type I,
while it is locked for type II. For example,
in case of (ii), $\lambda_0$ decreases up to $t=t_1^+$,
and is locked for a while till $t=t_1^-$, then is unlocked and
decreases again till $t=t_2^+$, and so on.
These locking-unlocking transitions occur many times as $t$ goes by.

\begin{figure}
\centering
\includegraphics*[width=\columnwidth]{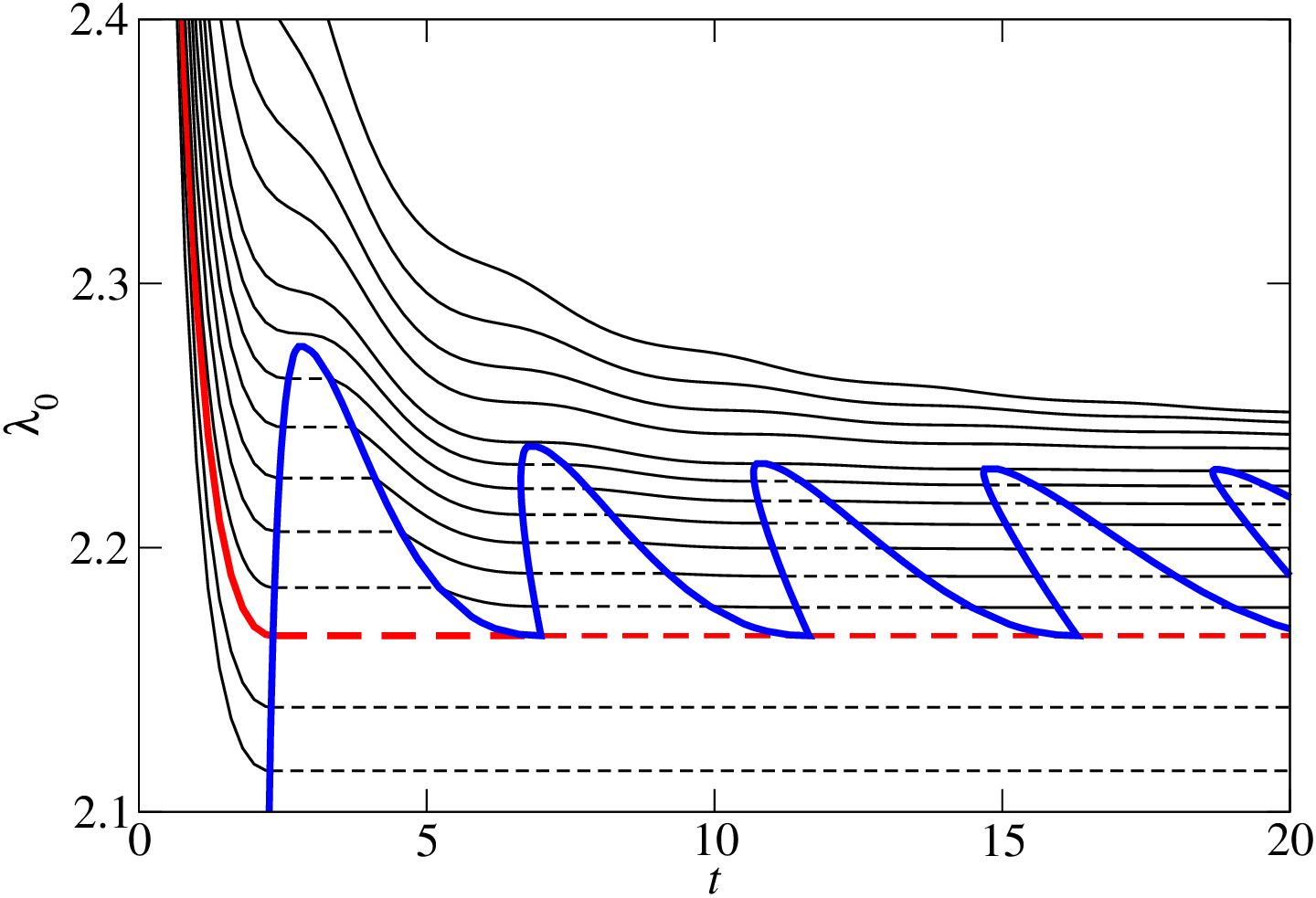}
\caption{(Color online) Plots of $\lambda_0$ vs $t$ for several values of $u$
at $\ndcoeff=3/10$.
The value of $u$ ranges from $0.55$~(top) to $0.76$~(bottom). The solid
curves correspond to the PDF of type I, while the dashed line corresponds to
the PDF of type II with the boundary drawn with the thick zigzag (blue) curve.
The thick (red) curve corresponds to the case with $u=u^*$.}
\label{fig2}
\end{figure}

\begin{figure}
\centering
\includegraphics*[width=\columnwidth]{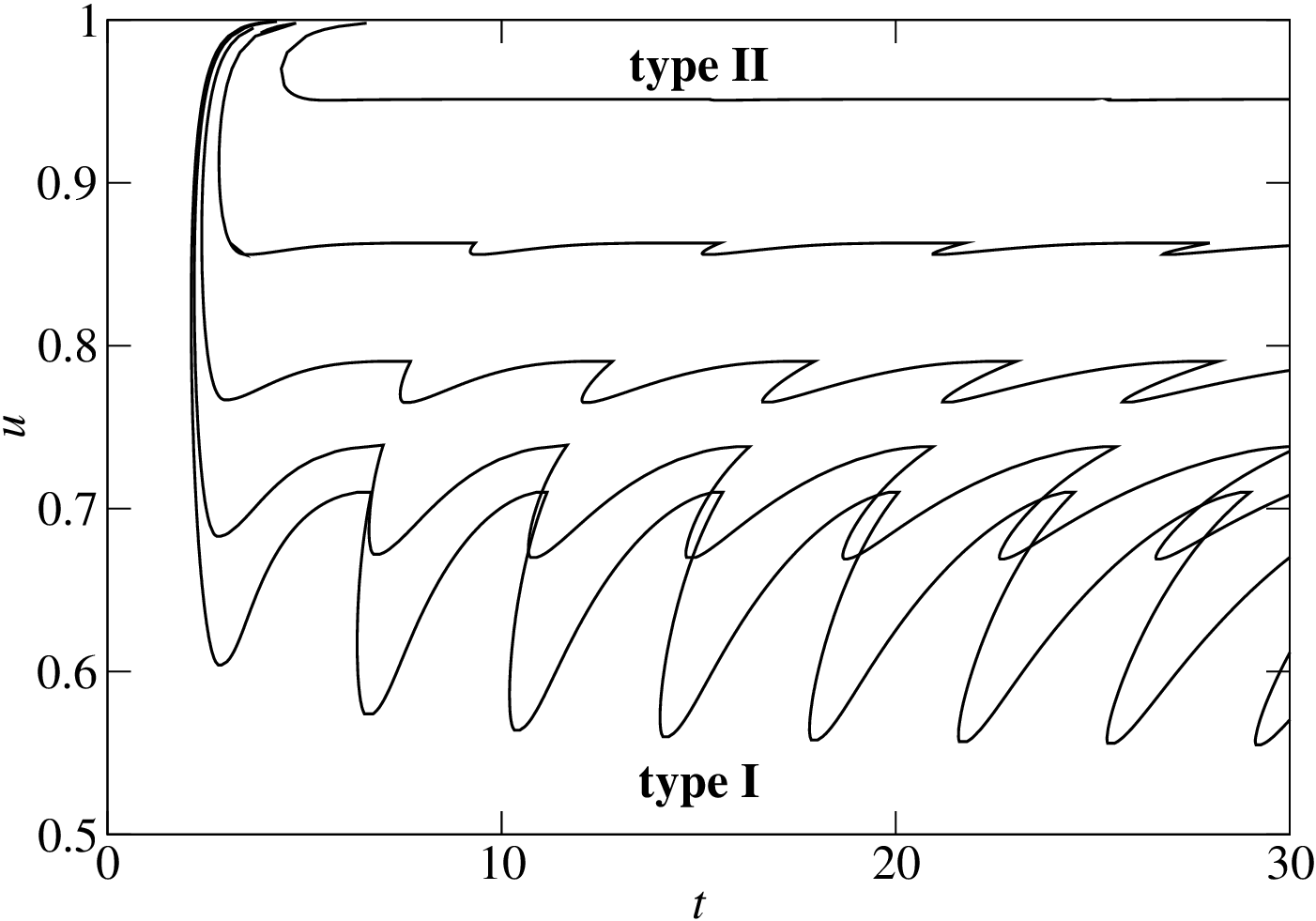}
\caption{Phase diagram in the $u$-$t$ plane for several values of
$\ndcoeff = 0.1$~(bottom), $0.3$, $0.5$, $0.7$, and $0.9$~(top).
The PDF belongs to type II in the region surrounded
by the phase boundary curve, and type I elsewhere except the line $u=0$
where the type-0 PDF is found.
For $\ndcoeff\ge 1$, the phase boundary disappears completely and no type II exists.
}
\label{fig3}
\end{figure}

For given values of $u$ and $\ndcoeff$, the time dependence of $\lambda_0 (t)$ can be
calculated numerically exactly from the roots of $\det\tilde{\mathsf A}(t;\lambda)=0$,
using the solution given in Eq.~(\ref{A_solution}). In Fig.~\ref{fig2}, we plot
$\lambda_0$ as function of $t$ for several values of $u$ at $\ndcoeff=3/10$.
When the anisotropy is weak~(small $u$),
the inverse characteristic work $\lambda_0(t)$ decreases smoothly with time~(upper curves in Fig.~\ref{fig2}).
With the intermediate anisotropy, it
decreases being locked for a while in multiple or infinite number of plateaus~(middle curves).
When the anisotropy is strong~(large $u$),
it decreases at small $t$ and then is locked to a constant value
forever~(lower curves).
Consequently, the PDF undergoes a single dynamic transition
for large values of $u$, infinitely many transitions for intermediate
values of $u$, and no transition for small $u$.
From the plots in Fig.~\ref{fig2}, one can construct the phase diagram
in the $u$-$t$ plane, which is drawn in Fig.~\ref{fig3}.
The phase diagram separates the regions with PDF of type I and type II.
When the anisotropy is weak~(strong), the system tends to display the PDF
of type I~(II). The phase boundary becomes complex in between, where multiple
dynamic transitions can be observed in the system with intermediate anisotropy.

In summary, our analytic result shows that such a simple LDS
exhibits surprisingly complex nonequilibrium fluctuations. It  raises
interesting questions for the mechanism of the dynamic transition.
The conservative part of the drift force
generates an anisotropic harmonic potential, which attracts the particle
toward the origin. The nonconservative part acts like a torque, which
drives the particle into a rotational motion. It is useful to
introduce the polar coordinate to focus on the rotational dynamics separately.
Then, the dynamics of the polar angle may be written down effectively as
$d\phi/dt = a \sin \phi + b +\xi$,
where the potential amplitude $a$
should be proportional to the anisotropy $\sigma$ and the constant driving force $b$ should
be proportional
to the strength of the nonconservative force $\ndcoeff$. This is the equation of motion
in a tilted periodic potential~\cite{Blickle07}.
For $a \ll b$ (small $\sigma$), the particle
has no time to relax in the potential well and drifts down into the steady state with a
constant velocity. So there is no extra time scale except for the relaxation into the
steady state. Thus, we expect no dynamic transitions and type-I PDF forever.
For $a\gg b$ (large $\sigma$), the particle sits at the potential well long enough and
fully relaxes inside the well. Then, it can hop to the neighboring
well due to the noise after a finite time, which can be determined by the noise strength.
This additional time scale exists and may set the transition time
$t_1^+$. As the particle relaxation in the first well is fully developed already,
there will be no additional time scale needed for successive hoppings. Thus, we expect
one dynamic transition from type-I to type-II PDF.
When $a\approx b$ (intermediate $\sigma$), it needs additional time scales for successive
hoppings with incomplete relaxations within each potential well, which leads to multiple dynamic transitions.
The above argument provides a plausible understanding of existence of multiple time scales,
but does not fully capture underlying mechanisms of dynamic locking-unlocking transitions.
Since the radial component also fluctuates in our model, there are more
possible complex routes to relax into the steady state. It is remarkable to find no smearing out
of sharp dynamic transitions with fully locking states.
We leave full intuitive physical understanding of these dynamic transitions for future works.

As can be seen in our analysis, the remarkable characteristics of multiple dynamic transitions
should not be a pathological property of some special systems. Hence, we expect that any high-dimensional
dynamical system driven by a nonconservative force should exhibit similar or more complex
multiple dynamic transitions in a reasonably large parameter space. It would be interesting
to observe these locking-unlocking features in experimental setups such as in
\cite{Hayashi10,Blickle07,Filliger07,Ciliberto13} and also in direct numerical simulations of
the Langevin equation. However, it demands extremely high-precision and time-dependent work PDF measurements
in rare-event regimes.

H.P.~thanks David Mukamel and Haye Hinrichsen for useful discussions.
This work was supported by the Basic Science Research Program through
the NRF Grant No.~2013R1A2A2A05006776(J.D.N.), 2013R1A1A2011079(C.K.), and 2013R1A1A2A10009722(H.P.).


\begin{thebibliography}{99}
\bibitem{exp} A nonconservative force matrix needs not be antisymmetric, in general
(see~\cite{Kwon11}).
\bibitem{comment} ${\mathsf F}_c$ may have a nonzero symmetric offdiagonal element. But
it can be set to be zero without loss of generality by a rotational coordinate transformation.
\bibitem{exp2} Detailed analysis of Type II can be found in~\cite{Kwon11}.
\end{thebibliography}
\end{document}